\documentclass[usenatbib]{mn2e}
\usepackage{apjfonts}
\usepackage{epsfig}
\usepackage{deluxetable}
\usepackage{amssymb}
\usepackage{amsmath}
\usepackage{fixltx2e} 

\newcommand{\scaleup}{}

\newcommand\plotone[1]
 {\centering \leavevmode \includegraphics[width={0.99\columnwidth}]{#1}}
\newcommand{\plotside}[1]
 {\centering \leavevmode \includegraphics[width={0.95\textwidth}]{#1}}
\newcommand{\acknowledgments}{\begin{small}\section*{Acknowledgments}\end{small}}
\newcommand\altaffilmark[1]{$^{#1}$}
\newcommand\altaffiltext[1]{$^{#1}$}
\newcommand{\etal}{et al.}

\newcommand{\msun}{M_{\sun}}

\title[The Mass Removed in Cores]{A Non-Parametric 
Estimate of Mass ``Scoured'' in Galaxy Cores}

\author[Hopkins \etal]{
\parbox[t]{\textwidth}{ 
Philip F.\ Hopkins\altaffilmark{1}\thanks{E-mail:phopkins@astro.berkeley.edu},
\&\ Lars Hernquist\altaffilmark{2} 
} 
\vspace*{6pt} \\
\altaffiltext{1}{Department of Astronomy, University of California 
Berkeley, Berkeley, CA 94720} \\
\altaffiltext{2}{Harvard-Smithsonian Center for Astrophysics, 60
Garden Street, Cambridge, MA 02138, USA} }

\date{Accepted to MNRAS, May, 2009}
\begin{document}
\maketitle
\label{firstpage}

\begin{abstract}

We present a simple estimate of the mass ``deficits'' in cored 
spheroids, as a function of galaxy mass and radius within the galaxy. 
Previous attempts to measure such deficits depended on 
fitting some functional form to the profile at large radii 
extrapolating inwards; this is sensitive to the assumed functional form 
and does not allow for variation in nuclear profile shapes. 
For example, we show that literally interpreting the residuals from a 
single/cored Sersic function fit as implied ``deficit'' can be misleading. 
Instead, we take advantage of larger data sets to directly construct 
stellar mass profiles of observed systems and measure the stellar mass 
enclosed in a series of physical radii ($M_{\ast}(<R)$), for samples of cusp and 
core spheroids at the same stellar mass. 
We show that there is a significant (model-independent) bimodality 
in this distribution of central 
structure for this sample at small radii.
We non-parametrically measure 
the median offset between core and cusp populations (the ``deficit'' 
$\Delta\,M_{\ast}(<R)$). We can then construct the scoured mass 
profile as a function of radius, without reference to any assumed 
functional form. 
The mass deficit rises in 
power-law fashion ($\Delta\,M_{\ast}(<R)\propto R^{1.3-1.8}$) from a significant 
but small mass at $R\lesssim10\,$pc, to asymptote to a maximum 
$\sim0.5-2\,M_{\rm BH}$ at $\sim100\,$pc, where $M_{\rm BH}$ is the
mass of the central, supermassive black hole hosted
by the spheroid.
At larger radii there is no statistically 
significant separation between populations; the upper limit to the cumulative 
scoured mass at $\sim$kpc is $\sim2-4\,M_{\rm BH}$. 
This does not depend strongly on stellar mass. 
The dispersion in $M_{\ast}(<R)$ appears larger 
in the core population, possibly reflecting the fact that core scouring 
increases the scatter in central profile shapes. We measure this broadening 
effect as a function of radius. The relatively low 
mass deficits inferred, and characteristic radii, are in good agreement with 
models of ``scouring'' from BH binary systems. 

\end{abstract}


\section{Introduction}
\label{sec:intro}

Massive ellipticals appear to exhibit central ``cores'' -- regions of 
constant or weakly divergent surface brightness -- as opposed 
to the power-law ``cusps'' -- more rapidly rising at small radii -- 
common in less massive ellipticals 
\citep{king78,young78,lauer85:cores,
kormendy85:profiles,kormendy:cores.review,lauer92,crane93,lauer:95,
ferrarese:type12,ferrarese:profiles,cote:virgo,cote:smooth.transition}. 
This bimodality in central surface brightness slopes 
\citep[][]{gebhardt96,lauer:bimodal.profiles}
participates in a larger, 
longer-recognized division of the elliptical population: the typical 
giant, core elliptical rotates slowly, and has ``boxy'' isophotal shapes 
supported by anisotropic velocity dispersions; less massive 
cusp ellipticals (and S0 galaxies and classical bulges) rotate 
more rapidly, have more isotropic velocity dispersions, and exhibit 
``disky'' isophotal shapes 
\citep{kormendy:bulge.rotation,
davies:faint.ell.kinematics,davis:85,jedrzejewski:87,
bender:88.shapes,bender89,bender:ell.kinematics,peletier:profiles,
kormendybender96,faber:ell.centers,
simien:kinematics.1,simien:kinematics.2,simien:kinematics.3,
emsellem:sauron.rotation.data,emsellem:sauron.rotation,
mcdermid:sauron.profiles,cappellari:anisotropy}.

These differences thus led naturally to the idea, developed in e.g.\ 
\citet[][]{faber:ell.centers,kormendy99,
quillen:00,rest:01,ravindranath:01,laine:03,lauer:centers,
lauer:bimodal.profiles,ferrarese:profiles,cote:smooth.transition,jk:profiles}, 
that disky, rapidly rotating cusp ellipticals are direct 
products of gas-rich (``wet'') mergers, whereas boxy, slowly rotating core 
ellipticals have been 
shaped by subsequent dissipationless (``dry'') re-mergers of 
two or more (initially cuspy gas-rich merger remnant) ellipticals. 
However, numerical experiments and simple phase-space 
arguments demonstrate that central light profile shapes are, to lowest order, 
preserved in dissipationless mergers without binary black holes 
\citep[e.g.][]{boylankolchin:mergers.fp,hopkins:cores}. It is therefore 
generally believed that the connection between the merger history 
of galaxies and their nuclear profile slope 
arises because of ``scouring'' by a binary black hole 
\citep[for a review, see][]{gualandrismerritt:scouring.review}. 
\citet{begelman:scouring} first pointed out that binary black holes 
coalescing in a dissipationless
galaxy merger could stall (i.e.\ are no longer efficiently 
transported to the center via dynamical friction) at radii $\sim$\,pc, 
larger than the radii at which gravitational radiation can efficiently 
dissipate energy and merge the binary -- the so-called 
``last parsec problem.'' They noted that significant gas content 
can provide a continuous source of drag and friction and 
solve this problem in gas-rich mergers, but that in ``dry'' mergers, the binary 
will remain stalled for some time, and will harden by scattering stars in the 
nucleus in three-body interactions. This will continue, flattening the 
nuclear slope, until a sufficient mass in stars 
($\sim M_{\rm BH}$, by simple scaling arguments) is ejected to merge 
the binary. 

It is therefore of particular interest to estimate the 
stellar mass which must be scattered to explain the slopes of cores, 
as a test of scouring models, a constraint on BH-BH merger rates 
and gravitational inspiral, and as a probe of the 
galaxy merger history. To date, there have been a few such 
calculations \citep[see][]{milosavljevic:core.mass,
ferrarese:profiles,lauer:bimodal.profiles,
lauer:massive.bhs,cote:smooth.transition,jk:profiles,kormendy:core.mass.deficits}; 
however, these have in some cases reached 
very different conclusions regarding the absolute mass 
``scoured,'' whether or not this scales with galaxy mass, 
and even whether or not there is any evidence for a bimodal 
population at all (as opposed to a smoothly scaling distribution of 
central profile gradients). 

Much of this ambiguity owes to the lack of 
an {\em a priori} physical model for the ``pre-scouring'' profile shape. 
All of the estimates above implicitly assume some specific functional 
form for the light ``scoured,'' where the parameters of this functional form 
are determined from fitting the observed light profile at larger radii 
and extrapolating inwards. For example, several of the above works 
fit either a ``cored'' Sersic model \citep{graham:core.sersic} or a 
Sersic model with some deficit inside a ``cutoff'' radius to the large-scale 
galaxy profile, and then compare the actual light profile with the 
inwards extrapolation of the single Sersic portion of the fit (implicitly assuming 
that the ``pre-scouring'' light profile followed the 
continuation of an identical Sersic profile to 
the outer, post-merger profile). 

However, 
there is no physical motivation for the Sersic model. 
It is well-known that 
there is considerable diversity in the central profile shapes within the 
cusp population -- any individual system could deviate widely from 
this assumption -- and moreover the assumption that profiles represent perfect 
Sersic continuations is, in fact, known to break down for the 
majority of non-core ellipticals \citep{faber:ell.centers,
hibbard.yun:excess.light,rj:profiles,
trujillo:sersic.fits,ferrarese:profiles,
cote:smooth.transition,jk:profiles,hopkins:cusps.ell}. 
For example, \citet{hopkins:cusps.mergers,hopkins:cusps.evol} show that 
a physically motivated decomposition into a two-component Sersic 
model more accurately approximates the central structure at $\sim$kpc scales 
in ellipticals. 
Adopting a slightly different assumed functional form for the profile fit and  
extrapolation leads to large systematic differences in the ``missing light'' properties. 
In addition, it is well-established 
that core and cusp ellipticals have different profile shapes at radii outside 
the ``core'' (see references above); even if all systems obeyed the 
assumed specific functional form, the use of the profile at large radii -- radii 
that have no ``information'' regarding the scouring process -- 
to anchor the fit could lead to significant differences between the 
true and inferred pre-scouring profiles \citep[see][]{milosavljevic:core.mass,
hopkins:cores}. 

Moreover, the scouring process is not expected to be identical 
from one system to the next. The efficiency and 
effects of scouring, in detail, are functions 
of the merger history \citep[e.g.\ number of mergers and distribution of merger 
mass ratios and orbits; see][]{quinlan:bh.binary.tang.orbit.bias,
merritt:mass.deficit,sesana:binary.bh.mergers}, as well as properties such as the 
triaxiality and stellar orbital anisotropy distribution as a function of radius
\citep{holley:triaxial.loss.cone,
berczik:triaxial.bh.mergers,
perets:massive.perturber.bh.mgr}; numerical simulations suggest that it may not be described 
by a simple functional form relative to the profile at larger radii 
\citep[see references above and][]{perets:binary.bh.loss.cone.filled.by.massive.perturbers,
zier:binary.bh.loss.cone.dynamics}.

In this paper, we therefore present a non-parametric estimator of the 
mass affected by BH scouring, as a function of radius. 
We construct stellar mass density profiles for a large sample of 
both cusp and core ellipticals, and use these to determine the stellar 
mass enclosed within different physical annuli. Considering this distribution in 
ellipticals of similar mass, we show that -- even without any reference to 
some ``cusp'' or ``core'' designation or profile fitting -- there is a 
robust bimodality in the distribution of enclosed mass at small radii. 
We consider non-parametric measures of the median offset between these 
bimodal peaks; equivalently, the median difference in enclosed mass between 
cusp and core ellipticals, as a function of radius. This allows us to construct -- albeit 
still with some assumptions -- the 
``scoured'' mass profile, which grows with radius until asymptoting to a maximum 
scoured mass $\approx1-2\,M_{\rm BH}$ near $\sim100\,$pc. 
We consider this as a function of galaxy mass, and with different 
methodologies, and show that it is robust. Unlike the approaches above, 
the methodologies here do not refer to specific profile fits, and account for the diversity of profile 
shapes within each population. Moreover this methodology allows us to 
actually measure the effects of scouring as a function of radius (rather than 
implicitly assuming them in the fitting); we show there is an approximate 
power-law behavior at small radii with an asymptotic effect on the mass 
profile at larger radii. 

We adopt a $\Omega_{\rm M}=0.3$, $\Omega_{\Lambda}=0.7$,
$H_{0}=70\,{\rm km\,s^{-1}\,Mpc^{-1}}$ cosmology, but note that this has
little effect on our conclusions.

\section{The Concern: Inference by Fitted Profile Shapes}
\label{sec:inferred}

\begin{figure}
    \centering
    \scaleup
    \plotone{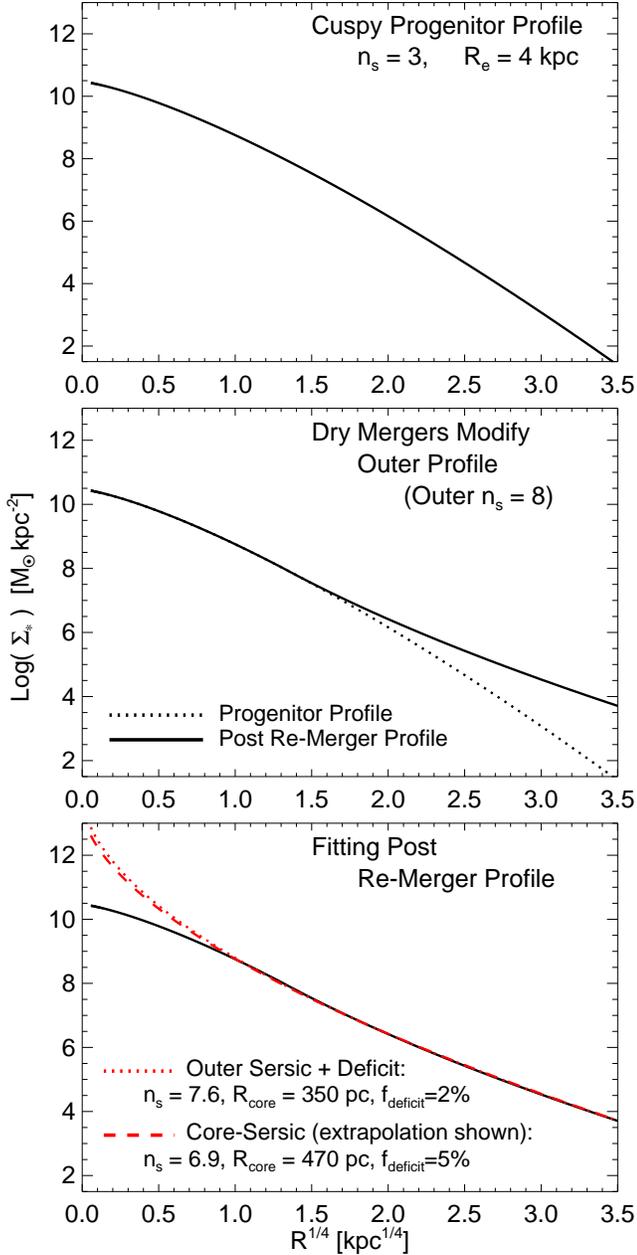}
    \caption{Illustration of caveats in estimating the ``missing'' mass in 
    core ellipticals via fitting some assumed functional form to the profile. 
    {\em Top:} An idealized profile of a typical cuspy progenitor elliptical: 
    here a perfect $n_{s}=3$ Sersic profile, with the given effective radius. 
    We plot surface stellar mass density versus radius. 
    {\em Middle:} One possible way in which subsequent re-mergers 
    and dry mergers will modify the profile, by building up an outer envelope of 
    low-density material (here modeled with an $n_{s}=8$ profile outside of 
    $R_{e}$). There is {\em no} change, however, in the central profile shape 
    -- in particular, no core ``scouring'' here. 
    {\em Bottom:} 
    The results of fitting these to assumed functional forms. First, a 
    single Sersic law, fit down from large radii until the variance 
    at a given radius becomes larger than the equivalent of $\Delta\mu=0.1\,$mag. 
    Comparing the inward extrapolation of this to the true profile gives an 
    apparent ``deficit'' (mass fraction $f_{\rm deficit}$) 
    inside of this radius ($R_{\rm core}$). 
    Second, a core-Sersic profile fit (fitted to the entire profile). We plot the extrapolation of 
    the outer Sersic portion of the fit (the best-fit full profile with ``core'' 
    is indistinguishable from the true profile); with the given 
    outer Sersic index, core break radius, and apparent core mass deficit. 
    \label{fig:cartoon}}
\end{figure}

To begin, Figure~\ref{fig:cartoon} illustrates one reason why caution is needed in 
the approach of fitting galaxy profiles. Begin with a simple toy-model 
galaxy light profile, a Sersic profile with index $n_{s}=3$. 
This is typical of ``cuspy,'' $\sim L_{\ast}$ ellipticals. 
Now imagine that the system undergoes some series of violent events, such that 
the profile at large radii is altered and an extended envelope is built up, but the 
profile at small radii is relatively unperturbed. 
This is a plausible scenario in the absence of core ``scouring'': in N-body and hydrodynamic 
simulations of ``dry'' mergers, even in major (1:1) mergers, the 
result to lowest order is that the profile shape in the central regions where the system is very dense 
tends to be conserved (stars there
conserve their specific binding energies; see, e.g.
\citealt{barnes:disk.halo.mergers,hernquist:bulgeless.mergers,
hernquist:bulge.mergers}), while some scattering or 
``splashing'' of stars to larger radii makes the tail or envelope of the distribution 
at large radii more extended \citep[see][]{boylankolchin:dry.mergers,hopkins:cores}. 

The effect is expected even more so if the envelope is built up not by a 
single very violent event, but by a series of minor mergers and/or mergers where the 
secondary is much less dense than the primary -- in which case it is shredded 
at large radii and contributes directly to the envelope, but has a 
completely negligible effect on the central densities 
\citep{gallagherostriker72,hernquist:phasespace,naab:size.evol.from.minor.mergers}. 
Recent evidence from comparison of 
high-redshift massive spheroids and their descendants today suggests, 
in fact, that the buildup of such an envelope while central properties are 
(relatively) weakly affected is probably the dominant evolutionary track for 
the most massive spheroids 
\citep{hopkins:density.galcores,bezanson:massive.gal.cores.evol,
hopkins:r.z.evol}.
We mock up such an expansion of the outer envelope by simply modifying the 
profile such that inside of the {\em original} $R_{e}$, it follows 
exactly the same functional form as before; but at larger radii, it obeys a 
Sersic law with $n_{s}=8$ (these choices ensure the profile is smooth through 
the transition). 

Now, fit the relic profile to a new Sersic function. 
We consider first a single Sersic function fit, following \citet{jk:profiles} -- 
we fit the profile at large radii first, and include more data points to 
smaller and smaller radii (re-fitting each time) until we reach some 
radius where the rms deviations about the inward extrapolation of the 
single Sersic law grow larger than $0.04$\,dex in $\log{\Sigma_{\ast}}$ 
(roughly equivalent to $\Delta\mu=0.1\,$mag in surface brightness). 
Similarly, we can consider the core-Sersic profile 
\begin{equation}
\Sigma = \Sigma'\,{\Bigl[}1+{\Bigl(}\frac{R_{\rm core}}{R}{\Bigr)}^{\alpha}{\Bigr]}^{\gamma/\alpha}\,\exp{{\Bigl[}
-b_{n}\,{\Bigl(}\frac{R^{\alpha}+R_{\rm core}^{\alpha}}{R_{\rm eff}^{\alpha}}{\Bigr)}^{1/(\alpha\,n)}{\Bigr]}}
\label{eqn:coresersic}
\end{equation}
which follows a Sersic 
function at large radius, with a break to some flatter power-law behavior at small 
radii \citep[the fitted $R_{\rm core}$;][]{graham:core.sersic}, fitted to all radii simultaneously. 
In either case, the resulting fit defines some outer Sersic profile, and some 
inner ``core'' region, inside of which the true profile falls below the inward 
extrapolation of the Sersic component. 

We show these extrapolations 
for both cases. They give similar best-fit $n_{s}\sim7$ (obviously 
similar to the input $n_{s}=8$ profile as $r\rightarrow\infty$). The inwards extrapolation 
of each of these fits rises to much larger surface densities than the 
true profile. Thus there is an apparent large ``core'' or mass deficit inside 
small radius. Here, the apparent deficit appears at $\sim400\,$pc, 
and has a mass or light fraction (defined by simply integrating the 
light under each curve versus the extrapolated Sersic component) of 
$\sim2-5\%$. 

\begin{figure}
    \centering
    \scaleup
    \plotone{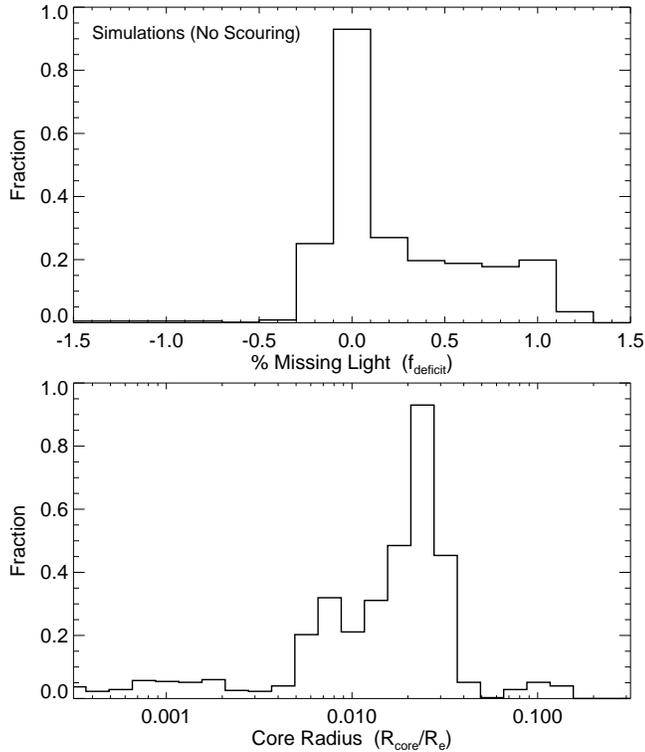}
    \caption{Distribution of ``mass deficit'' properties in $\sim30$
    hydrodynamic simulations of major dry merger remnants from \citet{hopkins:cores}. 
    The galaxies are initially in good agreement with profiles of observed cusp 
    ellipticals, and are merged with {\em no} scouring mechanism. Post-merger, 
    they are fitted to a core-Sersic profile, with the implied mass deficits and core 
    break radii determined as in Figure~\ref{fig:cartoon}. 
    {\em Top:} 
    Distribution of implied mass deficit percentages. 
    {\em Bottom:} 
    Distribution of implied core break radii ($R_{\rm core}$ in Equation~\ref{eqn:coresersic}, 
    in units of $R_{\rm eff}$). 
    Owing to the effects in Figure~\ref{fig:cartoon}, and simple variation/scatter in 
    the true functional form of galaxy profile shapes, apparent 
    non-trivial mass deficits and large break radii can arise.
    \label{fig:ml.correlations}}
\end{figure}

Of course, the inner regions of our mock profile are, by construction, identical 
to the progenitor -- there has been no scouring of any kind. 
Assuming that the inwards extrapolation of the fit from large radii represents 
the progenitor is therefore misleading. 
Although the example here is purely illustrative, such an effect 
appears in numerical simulations of dry galaxy-galaxy mergers, 
for the reasons above. 
In \citet{hopkins:cores}, the authors present a large suite of simulations of 
dry ``re-mergers'' of simulated elliptical galaxies, themselves the products of 
gas-rich mergers chosen specifically because of their very close agreement 
with the observed profile shapes of real observed cusp ellipticals 
\citep{hopkins:cusps.ell}. After a 1:1 dry re-merger of such systems 
(with no scouring mechanism included), the authors 
note that many of the profiles exhibit exactly the sort of changes described above 
(where the inner profile is approximately conserved, but outer profile ``puffed up''), and 
can be formally well-fit by a core-Sersic profile. This is despite the fact that 
there is no core scouring included or resolved in those simulations. 
Fitting each simulation 
with such a profile, we show the distribution of mass ``deficits'' (defined as 
above for the core-Sersic fit) and core break radii 
in Figure~\ref{fig:ml.correlations}. The characteristic 
deficits of $\sim0.1-1\%$ of $M_{\rm gal}$, and break radii of $\sim0.01-0.05\,R_{e}$, 
are characteristic of the above issue. 

The problem arises because of the nature of the Sersic profile, 
but could generally arise from any extrapolation from larger radii. 
Fitting such a profile implicitly couples the profile shape at large 
radii to that at small radii. For a Sersic profile, 
a more extended envelope at large radii corresponds to a larger 
index $n_{s}$. However, a larger $n_{s}$ {\em also} corresponds to a more 
steeply rising profile at small radii, as is clearly evident in Figure~\ref{fig:cartoon}. 
The effects on the inferred nuclear ``scoured'' mass can be large: an 
$n_{s}=2$ profile has just $0.06\%$ ($3\%$) of its mass within 
$1\%$ ($10\%$) of $R_{e}$, an $n_{s}=8$ profile has $2\%$ ($14\%$). 
Thus, any change in the profile shape at large radii will necessarily change the 
inferred ``progenitor'' profile at small radii. 
Moreover, observed core ellipticals tend to have significantly higher 
$n_{s}$ in their outer regions (more extended envelopes) than 
lower-mass cusp ellipticals \citep{trujillo:sersic.fits,ferrarese:profiles,jk:profiles}. 
This means that, assuming such an inward extrapolation, the assumed 
``progenitor'' small-scale profile in the cored systems is in fact more steep than that 
observed in any actual cusp elliptical. For example, the 
stellar mass profiles of simulated merger 
remnants in \citet{hopkins:cusps.ell,hopkins:cusps.fp} are not generically 
reproduced by a single Sersic function, but rather exhibit multi-component 
structure indicative of the mixed roles of dissipation and violent relaxation.

\section{Methodology: An Estimator of the Core Mass ``Deficit''}
\label{sec:method}

Given these concerns, we desire here to develop a 
more simplified but hopefully more robust check of mass deficits and 
core radii. 

We begin by considering the surface stellar mass 
density profiles $\Sigma(R)$ as a function of projected radius $R$ 
for ellipticals and S0 galaxies with {\em HST} imaging of their 
nuclei (and ground-based profile data at large radii, 
allowing accurate surface brightness profile measurements from 
$\lesssim 10$\,pc to $\sim50$\,kpc).
We compile observed surface brightness profiles from 
\citet{jk:profiles} and \citet{lauer:bimodal.profiles}; this consists of 
a total of $\sim180$ unique local 
ellipticals.\footnote{Note 
that although the 
composite (HST+ground-based) profiles 
were used in \citet{lauer:bimodal.profiles} to estimate effective radii, 
they were not 
actually shown in the paper.} 
The {\em HST} data allows robust classification of the cusp/core status 
in each system, discussed therein.
We determine stellar masses for each system in 
\citet{hopkins:cusps.ell}, based on the 
integrated observed 
optical luminosity in several bands, using the mass-to-light ratio 
calibrations as a function of color in \citet{bell:mfs} corrected 
to a \citet{chabrier:imf} IMF.\footnote{Varying the 
specific bands used to determine stellar masses makes little 
difference, and changing the IMF will systematically alter the stellar 
masses of all objects considered, but will not affect our comparisons.}
The isophotally averaged major axis profiles are measured in rest-frame 
optical; we convert to a stellar mass profile based on the measured 
total stellar masses and the assumption of a radius-independent 
stellar mass-to-light ratio. Most of the objects also have resolved 
color gradients -- we find that converting to a stellar mass profile
using the local color and a color-dependent $M_{\ast}/L$ 
makes no difference, as the color gradients are weak. Likewise, 
conversion to stellar mass profiles using stellar population gradients 
and comparison of profiles from different 
instruments and wavebands in these samples are discussed extensively 
in \citet{hopkins:cusps.ell,hopkins:cusps.fp}; the differences 
are much smaller than the variation between individual profiles, and 
do not affect our conclusions. 
The \citet{jk:profiles} sample is a volume-limited survey of the 
Virgo spheroid population; as such it includes few very massive 
galaxies ($M_{\ast}>3\times10^{11}\,\msun$). The \citet{lauer:bimodal.profiles} 
galaxies are chosen to be characteristic of massive ellipticals 
in the local Universe, including more massive systems up to 
a couple $10^{12}\,\msun$. At the masses of interest, both are 
representative of the distribution of spheroid sizes, velocity dispersions, 
and profile shapes in the local SDSS galaxy 
sample \citep[see e.g.][]{hyde:stellar.mass.fp}.

\begin{figure*}
    \centering
    \scaleup
    \plotside{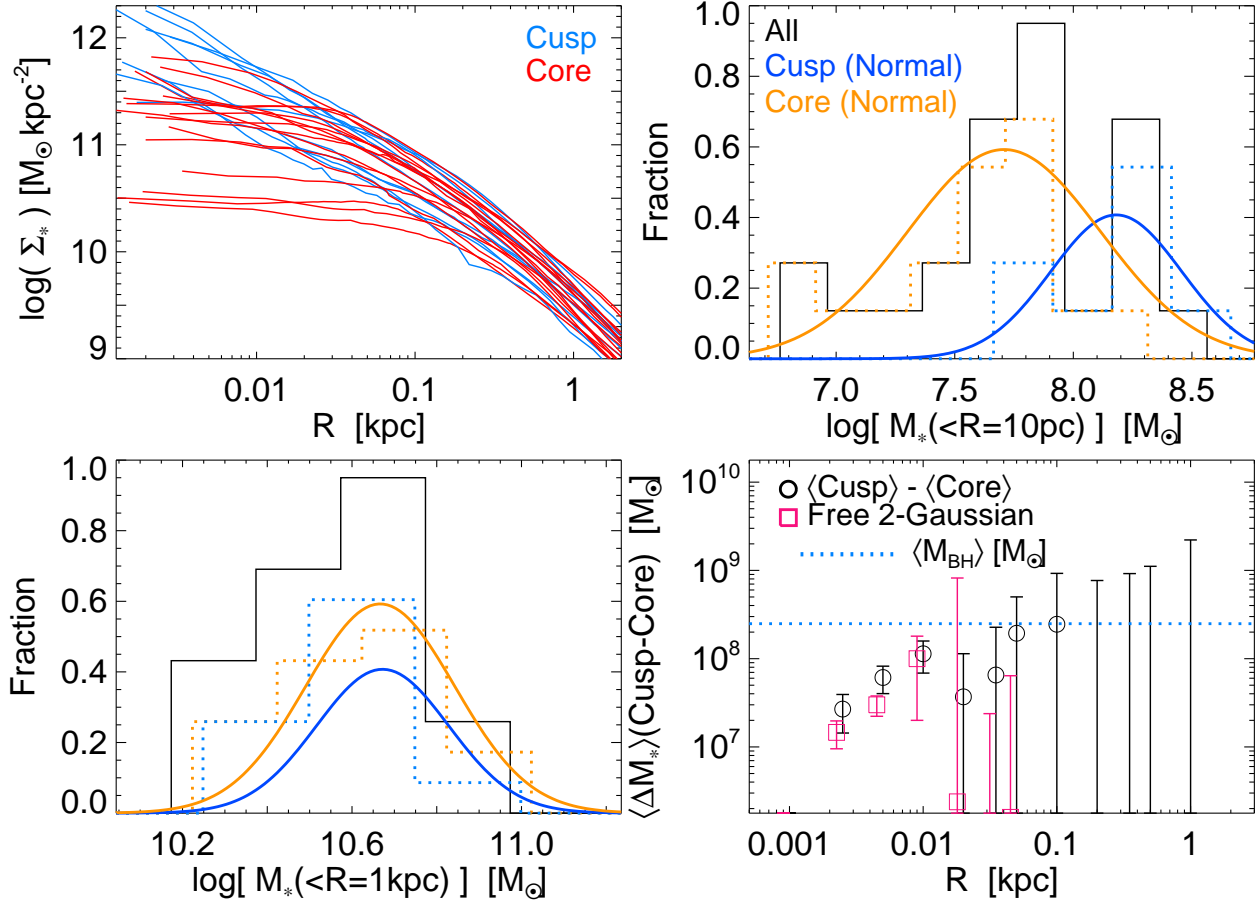}
    \caption{{\em Top Left:} Stellar mass surface density profiles of 
    spheroids in a narrow stellar mass range $M_{\ast}=10^{11-11.5}\,\msun$. 
    Cusp and core ellipticals are separately noted. 
    {\em Top Right:} Histogram of the enclosed mass $M_{\ast}(<R)$ inside 
    some radius, determined from the surface density profiles, at $R=10\,$pc. 
    There is a clear bimodality. Gaussian fits to the cusp and core populations separately 
    are shown (histogram of each in dotted lines). 
    {\em Bottom Left:} Same, at $1\,$kpc. 
    {\em Bottom Right:} Difference $\Delta\,M_{\ast}$ 
    between the median enclosed mass $M_{\ast}(<R)$ at each $R$, comparing the 
    cusp and core populations (black circles). Points represent detection of a statistically 
    significant offset in the median $M_{\ast}(<R)$ between the populations, with the 
    absolute mass noted. Error bars extending below the plotted 
    limits should be considered upper limits; in this case the observations are 
    consistent with no difference. Magenta squares fit the histogram to a double-Gaussian, 
    i.e.\ assume no prior cusp/core designation (as such the uncertainties are larger); 
    they are plotted only where a double Gaussian is favored at any significance over a 
    single Gaussian. Compare the average BH mass in galaxies of this stellar mass range. 
    There is a clear bimodality at small scales, where $\Delta\,M_{\ast}(<R)$ grows 
    with $R$ until appearing to asymptote to a maximum $\sim M_{\rm BH}$ at 
    $R\sim100\,$pc. 
    \label{fig:demo}}
\end{figure*}

Figure~\ref{fig:demo} illustrates the methodology. First, consider a narrow 
range in total galaxy stellar mass (here $M_{\ast}=10^{11.0-11.5}\,\msun$), 
in which there are both cusp and core elliptical populations. 
Ideally, we would also consider a specific range in black hole mass, 
but direct BH mass measurements are available for only a small subset of these 
objects (too few for the statistics here); as such, we adopt the observed 
BH-bulge stellar mass relation from \citet{haringrix} to estimate the corresponding 
BH masses in these hosts (more on this below). 
We show a direct comparison of the surface stellar mass density profiles 
of these systems. We separately denote the cusp and core ellipticals.\footnote{
For consistency, we adopt the cusp/core designations in 
the source samples from \citet{jk:profiles} and \citet{lauer:bimodal.profiles}, 
based on their logarithmic nuclear profile slopes at the smallest radii observed. 
However, other classifications are identical for all but a few objects; 
adopting those in \citet{ferrarese:profiles} or 
\citet{cote:smooth.transition} makes no difference. 
}
We plot results down to the minimum quoted radii from the sources above 
(the minimum radii at which their seeing-deconvolved profiles are considered reliable) -- essentially 
the HST resolution limit. We stress that we will use these points in all of our analysis below, so some 
caution is needed in considering how robust the conclusions we quote below are at the 
smallest radii. In general, resolution limits are such that below $\sim10\,$pc, the sample size with 
any resolved points drops dramatically. In the more massive galaxies, the limits are larger; 
the most massive systems generally are limited to $>30-50\,$pc resolution. If we 
simply interpolate the nuclear profiles inwards based on their slope at the minimum resolved 
radii, our results remain similar down to arbitrarily small radius, but we are specifically 
attempting to avoid such assumptions about the ``true'' profile shape, so in general one should 
consider the smallest-radii points in each mass bin to be fairly uncertain. 

Bearing those caveats in mind, at some fixed radius $R$, we can consider the distribution of 
stellar mass enclosed within that radius, i.e.\ 
\begin{equation}
M_{\ast}(<R) = \int_{0}^{R}\Sigma(R^{\prime})\,2\pi\,R^{\prime}\,{\rm d}R^{\prime}\ . 
\end{equation}
Note that this is for the projected surface density $\Sigma$. 
We can also apply an inverse Abel transform to de-convolve the (circularized) 
projected density profile to recover the three-dimensional density, 
\begin{equation}
\rho(r) = -\frac{1}{\pi}\,\int^{\infty}_{r}\frac{{\rm d}\Sigma}{{\rm d}R}\,\frac{{\rm d}R}{\sqrt{R^{2}-r^{2}}}
\end{equation}
\citep{bracewell:book} and then integrate to determine  
$M_{\ast}(<r)$ as the mass within some three-dimensional spherical 
radius. Because $M_{\ast}$ is an integral quantity, this makes little 
difference; in what follows we prefer the use of $M_{\ast}(<R)$ as it is 
more numerically stable (the gain in accuracy in de-projecting to a three-dimensional 
profile is offset by the fact that the Abel transform required in that de-projection is 
very sensitive to small uncertainties in the measured profile; as such a careful 
error analysis shows no improvement in the constraints). 
Figure~\ref{fig:demo} shows this for the systems in the given 
total mass range, at $R=10\,$pc and $R=1\,$kpc.

We first consider the cusp and core systems separately; Figure~\ref{fig:demo} 
plots the best-fit Gaussian to each distribution (i.e.\ Gaussian with the 
same median and $1\,\sigma$ dispersion), and the cumulative 
distribution (i.e.\ histogram of all sources, without reference to their 
cusp/core status). 
At small radii $\sim10\,$pc, there is a clear difference between the 
distribution of cusp and core enclosed masses -- significant 
at $\sim 5\,\sigma$. This is clear directly in the total histogram -- there is a 
striking bimodality in the enclosed mass. 
Comparing a double versus single normal distribution, we find
that a distribution with two distinct peaks is preferred at $>4\,\sigma$. 
Other tests yield similar results; 
for example, the ``dip'' test of \citet{hartigan:dip.bimodality.test} favors 
bimodality at $\sim3-4\,\sigma$ without reference to an assumption of 
normality in the two-peaked distribution, and can be used to 
non-parametrically 
quantify (albeit with larger uncertainties) the peak separation, yielding 
similar results. 
This should not be surprising; the statistical 
significance of the bimodality in central galaxy densities (and their slopes) 
has been the subject of a number of works \citep[see][]{ferrarese:type12,
gebhardt96,faber:ell.centers,lauer:bimodal.profiles}, 
and is well established in at least {\em this} sample of chosen cusp-core 
galaxies; whether it is smoothed out in strict volume-limited samples 
(the degree of ``dichotomy'' or 
the uniformity of the transition) is however a question still being debated 
\citep[see][]{cote:smooth.transition}. On the other hand, however, 
at $1\,$kpc, the distribution is clearly unimodal, and there is no 
significant difference between cusp and core profiles. 

Given this bimodality, we can quantify the difference, on average, between the 
mass enclosed at small $R$ in the populations. Specifically, we can take the
absolute difference between the {\em median} $M_{\ast}(<R)$ for cusp and core ellipticals, 
with the appropriate uncertainty reflecting both the uncertainty in each median and 
the weighted uncertainty in the difference. This is non-parametric, and allows us to 
directly compare cusp and core profiles at a given mass without reference to some 
model for what the 
profile ``should be.'' Moreover, the large scatter in $M_{\ast}(<R)$, reflecting the scatter 
in profile shapes, is properly accounted for -- we can robustly quantify the 
{\em median} mass difference in the populations.  And it also has the advantage that 
we can robustly define an error bar on the mass ``deficit.'' 

Figure~\ref{fig:demo} shows the result; on this scale, the two populations are offset 
by a median $\Delta M_{\ast}(<R=10\,{\rm pc}) = 1.1^{+0.5}_{-0.4}\times 10^{8}\,\msun$. 
Compare this to the average BH mass expected and observed 
in spheroids of this total stellar mass range, 
$\langle M_{\rm BH} \rangle \approx 2.5\times10^{8}\,\msun$
\citep[using $M_{\rm BH}\approx0.0014\,M_{\ast}$, from][]{haringrix}. 

We can be even more conservative, and make no reference to the prior designation 
of systems as ``cusp'' or ``core.'' Instead, we simply fit the cumulative distribution 
of $M_{\ast}(<R)$ to a double-peaked (double Gaussian) distribution 
(equivalently, we non-parametrically 
measure the double-peak separation from other statistical tests as discussed above). 
In this case, the ``separation'' is only well-defined if the double-peaked fit 
provides a significantly better fit than a single-peaked fit; if a double-Gaussian 
fit is significantly better 
(with the attendant 6 degrees of freedom instead of 3), i.e.\ results in both 
$\Delta\chi^{2}>1$ and improved $\chi^{2}/\nu$, then we quantify the 
resulting difference in the median between the two fitted peaks and the formal 
statistical uncertainty associated. Unfortunately, because the number of objects 
in any mass bin is small, we find that freeing {\em all} the parameters of the fit 
yields too many degrees of freedom for any but marginally significant results 
(although the results, averaging over all mass bins, reach $\sim3\,\sigma$ 
significance and support our conclusions from the prior-constrained comparison 
of cusp and core populations above). 

We therefore show results 
fixing the relative fraction of the population in each Gaussian to the fraction of 
cusp or core ellipticals (we have also fixed the relative widths, rather than the 
fractional population, and find similar results). If the division between cusp and 
core populations did not reflect some physical bimodality, this measure would give a 
null result (no signal for statistically significant difference) at all radii; if a large 
fraction of objects were somehow misclassified, or if an ``intermediate'' class exists 
(i.e.\ a smooth transition from cusp to core), 
this method will allow for that (even with our fixing of either the Gaussian widths or 
relative normalizations), and should still robustly quantify the difference between 
bimodal peaks in the population. We find similar results to the 
case above, but with larger 
uncertainties. 

Figure~\ref{fig:demo} continues to show these 
fit results as a function of radius, from 
$\sim3\,$pc to $\sim 10\,$kpc. Again, we recall the cautions stated above 
regarding the effects of resolution and seeing limits at the smallest radii. 
At $\gtrsim 100\,$pc, however, 
the distribution of enclosed masses is clearly unimodal. 
Specifically, at $\sim1\,$kpc (where we show the full distribution) for example, 
fitting directly the cumulative distribution there is no preference for a 
doubly-peaked underlying distribution.\footnote{
At these radii, caution is needed 
adopting the fully freed bimodal fitting approach -- if the distribution is 
slightly non-Gaussian (has some skewness), and one fits a double Gaussian, 
it is possible to obtain a statistically significant ``offset'' that is clearly unrelated 
to the cusp/core populations. Some constraint like that adopted here is necessary, 
or more arbitrary functional forms (still unimodal) must be fit in combination, 
to prevent unphysical results. Given these uncertainties, properly marginalizing 
over the fitting parameters is particularly important to obtaining 
valid error estimates. 
} 
Even considering the cusp and 
core populations separately,  
there is no statistically significant difference between their $M_{\ast}(<R)$ 
distributions on this scale. 
Formally, we find $\Delta M_{\ast} = -3\times 10^{7}\,\msun$ ($+1.7\times10^{9}\,\msun$) 
($-2.0\times 10^{8}\,\msun$); in other words, the distributions favor no difference 
between the two populations, and we can set a $90\%$ upper limit on the core 
``deficit'' at this radius of $\approx 4\,M_{\rm BH}$.


\begin{figure*}
    \centering
    \scaleup
    \plotside{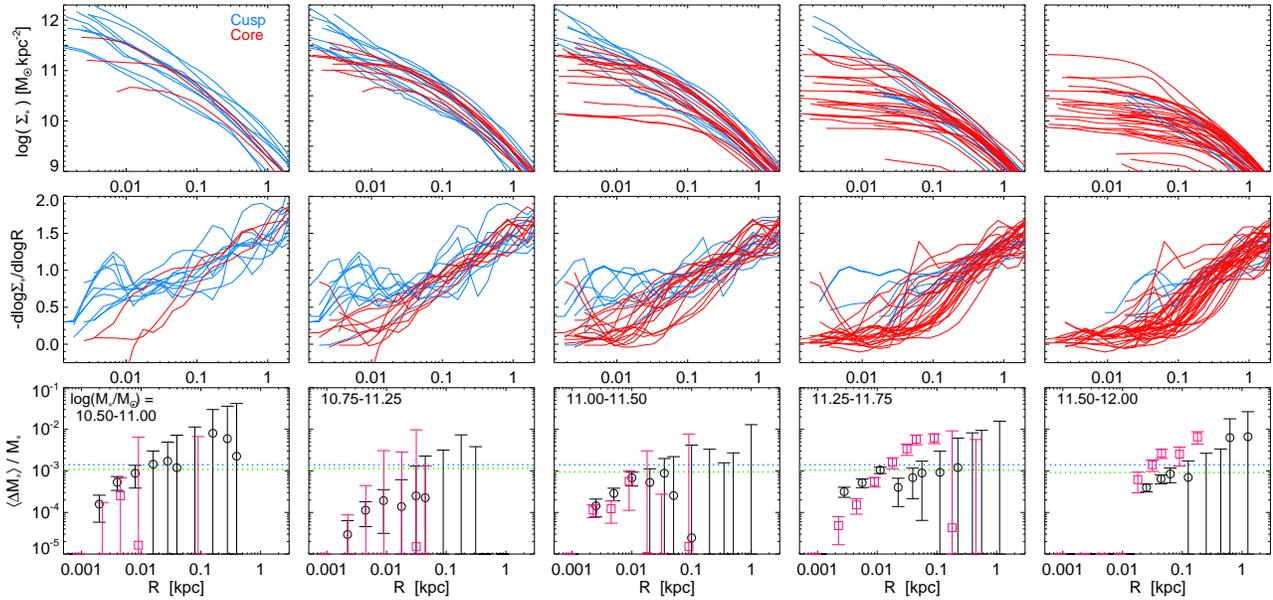}
    \caption{As Figure~\ref{fig:demo}, for spheroids 
    in different stellar mass intervals. 
    {\em Top:} Mass profiles. 
    {\em Middle:} Logarithmic slope of the mass profile, which highlights 
    the degree of ``flattening'' in the central regions. 
    {\em Bottom:} Estimated ``deficit'' as in Figure~\ref{fig:demo}, as a function of radius.     
    To compare systems with different total 
    masses, we plot the fractional mass difference enclosed, i.e.\ 
    $\Delta\,M_{\ast}(<R) / M_{\ast}$ (normalizing each profile by its stellar mass; 
    because of this, the plotted results for the $\sim10^{11-11.5}\,\msun$ bin 
    are not exactly identical to those in Figure~\ref{fig:demo}, though they are very 
    similar). Dotted lines show the expected $\langle M_{\rm BH} \rangle$ for the 
    median galaxy stellar mass in the bin and the BH-host mass relation of 
    \citet{haringrix}, or the median velocity dispersion $\sigma$ and BH-$\sigma$ 
    relation of \citet{tremaine:msigma} (blue and green, respectively). 
    In each mass bin, similar trends are seen: a significant bimodality at 
    small radii with $\Delta\,M_{\ast}(<R)$ growing to a maximum detection 
    at $\sim0.5-2\,M_{\rm BH}$ at $R\sim 100\,$pc; in at least several mass bins, 
    $90\%$ upper limits of $\Delta\,M_{\ast}<2-4\,M_{\rm BH}$ are 
    present out to $\sim 0.5$\,kpc. 
    \label{fig:fits}}
\end{figure*}

In Figure~\ref{fig:fits}, we generalize this to several mass intervals, 
from $\sim10^{11}-10^{12}\,\msun$. 
In each of several mass intervals, we again show the 
mass profiles of the observed cusp and core systems at small radii. 
We also show the logarithmic slope of the mass profile, which highlights the 
flattening in the cusp systems (discussed further below). 
And we show the median $\Delta M_{\ast}(<R)$ as a function of radius $R$, 
determined using the same methodology as in Figure~\ref{fig:demo}. 
To put the different mass intervals on the same footing, we 
show the mass fraction $M_{\ast}(<R)/ M_{\ast}(<R\rightarrow\infty)$. 
In these units, the average BH mass implied by the \citet{haringrix} BH-host 
galaxy mass correlation is a constant $\approx 0.0014$. 
That is not, of course, the only determination of this relation \citep[compare][]{magorrian,marconihunt}, 
nor is it the unique predictor of BH mass: velocity dispersion \citep{FM00,Gebhardt00}, 
host galaxy binding energy or potential well depth \citep{hopkins:bhfp.obs,hopkins:bhfp.theory,
aller:mbh.esph}, and profile shape \citep{graham:concentration} 
are possible alternatives. However, given the relatively limited mass range shown, these predictors 
do not differ very dramatically over this range. Specifically, we also show the 
median $M_{\rm BH}/M_{\ast}$ expected in each bin using the $M_{\rm BH}-\sigma$ 
relation together with the measured $\sigma$ of each galaxy in the subsample. 
At the highest masses, the median expectation from $M_{\rm BH}-\sigma$ is a factor of 
$\sim2$ lower than that from the $M_{\rm BH}-M_{\rm bulge}$ relation, but 
this is where our error bars become quite large (owing to the dearth of cusp ellipticals 
at these masses), so our comparison cannot discriminate at any significance whether 
there is a better correlation between the implied deficits and the BH-host 
mass implied BH masses or BH-$\sigma$ implied masses.
It should however be born in mind that the two estimators will diverge for the 
most massive systems, near the upper limit of our sample 
\citep[by $M_{\ast}=10^{12}\,\msun$ they may differ by a factor of several,][but most of the 
objects in our most massive bin are concentrated 
near the low-mass range of the interval]{lauer:massive.bhs}. 

In each case in Figure~\ref{fig:fits}, a similar trend is seen. At very small radii $\lesssim10\,$pc, 
there is a significant bimodality in the population, but the amount of 
mass contained in this radius is small. As such, the absolute mass 
difference between the cusp/core populations grows to larger and 
larger radii (very roughly with a power law-index 
$\Delta M_{\ast} \propto R^{1.3-1.8}$). However, between $\sim 10-100\,$pc, 
this appears to asymptote to a maximum of $\Delta M_{\ast}\sim1-2\,M_{\rm BH}$. 
Comparing each panel with the appropriate implied $\langle M_{\rm BH} \rangle$ 
from either the BH-host mass or BH-$\sigma$ relation, $\Delta M_{\ast}$ 
appears similar in units of $M_{\ast}$. 
At larger radii $R>100\,$pc, there is no statistically significant bimodality 
remaining. Formally, the difference $\Delta M_{\ast}\rightarrow 0$ at these 
radii; if this were strictly the case, it would imply that scoured mass is simply ``kicked'' 
to slightly larger radii (not expelled from the galaxy or moved to very large 
radii). However, the error bars are still large -- it is very possible that the mass is 
removed entirely, and so the mass difference $\Delta M_{\ast}$ remains constant 
at $\sim$a couple $M_{\rm BH}$ at all $R$ (of course, this will become less 
significant relative to the total enclosed mass at larger $R$). 
But the difference does not appear to grow; the 
$\sim 90\%$ upper limit in at least some cases remains at or below 
$\sim4\,M_{\rm BH}$ at radii $\sim 0.5-2\,$kpc. 

Of course, the comparison of profile shapes makes it immediately clear that the uncertainties 
at both low and high masses are considerable, owing to the lack of 
core and cusp spheroids, respectively. Note that the total mass range in Figure~\ref{fig:fits} 
is relatively small, and the individual bins in mass are relatively wide. This is necessary for the 
statistics needed, and because cusp galaxies become progressively more rare at high 
masses as cores are more rare at low masses. But together with the limited sample sizes 
observed at sufficiently high resolution, it means that the difference between the median 
mass of the cored spheroids in our lowest-mass bin and that of the cusp spheroids in our 
highest-mass bin is only a factor of $\approx7$. 
And recall that the apparent cusps or cores at the mass ``extremes'' may also represent 
unusual growth histories rather than typical progenitors/descendants. 
The effect of resolution limits is also increasingly evident in the highest-mass interval, 
where the innermost point of any comparison is at $\approx 20-30\,$pc 
(with the best statistics at $\sim50-100\,$pc). 
Thus our results at the extremes should be taken with appropriate caution. 

Figure~\ref{fig:fits} also shows the logarithmic derivative of the mass profile, 
$-d\log{\Sigma_{\ast}}/d\log{R}$, as a function of radius. This highlights the 
range of behavior in the central regions of the observed systems. 
It also reaffirms a number of our results. At large radii, it is clear that it is not just 
the absolute values of $\Sigma_{\ast}$ that are similar in all 
systems in each bin, but also their slopes -- this is where 
we do not see any statistically significant separation. 
At small radii, however, we see what is expected -- the galaxies identified as 
``cusp'' systems generally maintain slopes in the range 
$-1\lesssim d\log{\Sigma_{\ast}}/d\log{R} \lesssim -0.5$, down to the 
resolution limits, whereas those identified as cores fall closer to 
flat slopes. Of course, there are also intermediate systems in between.
Most interesting, the radii at which we see $\Delta M_{\ast}$ appear 
to asymptote (or above which do not see a statistically significant difference) 
appear to roughly correspond to the radii above which the 
core populations reach slopes steeper than $d\log{\Sigma_{\ast}}/d\log{R}\lesssim -0.5$. 
In the most massive core systems (although it is more ambiguous in the 
less massive core systems) this is also typically within a factor $\sim2$ of the 
radius where the absolute value of the 
second derivative of the profile slope is maximized (if it is estimated with 
some smoothing kernel; we do not show the numerical second derivatives here because they are 
very noisy). 

This is interesting because such a threshold has been proposed and 
used as an estimator of where ``scouring'' action has taken place -- if valid, 
it would allow an assessment of the scouring effects in {\em individual} galaxies, 
without the limitation of large samples required here 
\citep[see][]{gebhardt96,lauer:massive.bhs}. 
Of course, we are expressly attempting to avoid making assumptions about the 
functional form of the pre-scouring profile, so a comprehensive comparison of these 
results to those here is outside the scope of this paper, but the comparison 
here may lend some motivation to those approaches, if the threshold slope values 
are taken from comparison similar to that above. More rigorously, we can ask: 
if we were to adopt some parametric extrapolation of the profile shape on a per-galaxy 
basis, what would provide the best match to our results? 
Consider each mass interval above. For each core system, we could assume (with the 
dangers outlined above inherent in this process) that the pre-scouring profile 
followed a power-law at small radii with approximately constant slope at 
some $d\log{\Sigma_{\ast}}/d\log{R}=\eta_{\rm crit}$; we simply would estimate the 
``pre-scouring'' profile by extrapolating the observed profile with $\eta_{\rm crit}$ at 
radii below that where the observed profile slope falls below $\eta_{\rm crit}$. 
Averaged over the core population in each mass interval, this will give some 
average $\Delta M_{\ast}(<R)$ implied -- and we can ask what value of $\eta_{\rm crit}$ 
minimizes the difference between this estimator and our non-parametric estimator. 
Performing this exercise for the mass intervals in Figure~\ref{fig:fits}, we find 
that in all mass bins, the difference between our estimator and the distribution 
of deficits recovered by this method is minimized for 
threshold slopes in the range $-0.5\lesssim\eta_{\rm crit}\lesssim-0.8$ 
(specifically $-0.75,\,-0.52,\,-0.81,\,-0.63,\,-0.60$, from 
least to most massive, with respect to the cusp vs.\ core estimator; the 
error bars are sufficiently large in the 2-Gaussian estimator that there is 
not a strong discriminant in this range). 
Considering all bins simultaneously, the best-fit value is $\eta_{\rm crit}\approx-0.62$, 
for this sample.

\begin{figure}
    \centering
    \scaleup
    \plotone{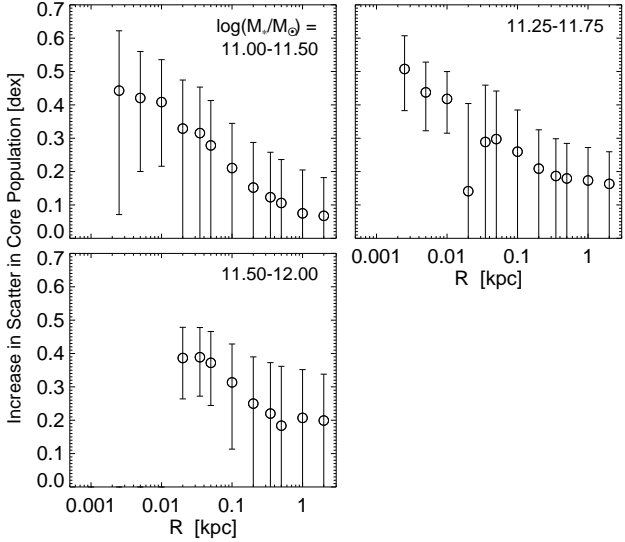}
    \caption{Difference in the logarithmic dispersion in $M_{\ast}(<R)$ 
    in cusp and core populations ($\sigma[\log{\{ M_{\ast}(<R) \}}]$, 
    for each population, subtracted in quadrature). 
    Where this is $>0$, it implies a significant broadening of the 
    dispersion in $M_{\ast}(<R)$ by the plotted dispersion. 
    Style is as Figure~\ref{fig:fits}; we exclude the two low-mass bins 
    as they show no statistically significant detection at any radius. 
    We find no statistically significant cases where the dispersion in core systems
    is more narrow than that in cusp systems. The observations tentatively 
    show evidence for non-trivial variation in scouring processes at small radii. 
    \label{fig:sigma}}
\end{figure}

At each radius, we have not just the median $M_{\ast}(<R)$ for 
each of the cusp and core populations, but also the dispersion 
(roughly lognormal) in $M_{\ast}(<R)$, $\sigma[\log{\{ M_{\ast}(<R) \}}]$.
At several radii in the mass bins considered, this dispersion 
appears to be significantly larger for the core population, 
as compared to the cusp population. In contrast, there are no cases 
of statistical significance where the cusp population has larger 
dispersion. This might be anticipated, given the arguments in \S~\ref{sec:intro}; 
the scouring process is, in detail, sensitive to quantities such as the 
merger mass ratio, binary orbital parameters, and triaxiality of the host, 
that will vary even at fixed stellar mass and fixed host mass profile shape. 
As such, it is plausible that scouring will broaden the distribution of 
mass profile shapes and correspondingly $M_{\ast}(<R)$ at 
small radii, relative to the distribution present in the progenitor population. 
If both $\sigma[\log{\{ M_{\ast}(<R) \}}]$ for core and cusp 
populations can be determined, this additional broadening or dispersion  
term should be given by their difference (subtracted in quadrature). 
Figure~\ref{fig:sigma} shows this. Because this requires going to 
higher-order moments than the median $M_{\ast}(<R)$, the statistics 
are correspondingly more limited -- we only find a statistically significant 
result at a couple of radii and masses. But in these cases, the trend is 
interesting. Reflecting what is seen in the absolute mass differences, 
the ``additional'' dispersion term appears to grow at small radii 
\citep[although it should be stressed that the dispersion in {\em both} 
populations grows at small radius; see][]{hopkins:msigma.scatter}. 
At $\sim1-10\,$pc, scattering processes may lead to fractional 
dispersion in $\Delta M_{\ast}(<R)$ at the $\sim0.5$\,dex level. 
In absolute terms, this amounts to dispersion, at these radii, of $\sim M_{\rm BH}$ 
in the absolute mass differences. At larger radii, the fractional dispersion 
falls (as it must, if the absolute mass scattered does not significantly exceed 
$\sim M_{\rm BH}$).

\section{Discussion}
\label{sec:discussion}

We illustrate important systematic uncertainties in the 
mechanism of inferring the ``mass deficits'' at the centers of cored 
spheroids via fitting of prior assumed functional forms to the mass 
profile (the ``core'' in the ``core-Sersic'' profile), and 
show that these can in principle lead to large artificial mass deficits 
and large apparent core radii. Specifically, we show that the assumption 
that the ``un-scoured'' progenitor profile matches the inwards extrapolation of a 
single Sersic fit to the galaxy profile at larger radii can be problematic, and 
is often misleading when applied to simulated galaxy profiles. 

To avoid these uncertainties, we therefore 
define a simple, non-parametric measure to compare the 
stellar mass difference in the center of cored versus cusp spheroids. 
Essentially, this amounts to a comparison of the enclosed mass 
profiles $M_{\ast}(<R)$, quantifying the significance of bimodality in 
this distribution and, given that, the separation between bimodal peaks. 
This does not rely on any prior assumption 
regarding the functional form of the ``true'' pre-scouring profile (only the 
concept that core ellipticals descend from cusp ellipticals similar 
to those present today); nor does 
the inferred mass deficit depend on the profile behavior at larger radii. 
It also allows us to account for the non-trivial dispersion in central profile 
shapes within the cusp and core populations. 

We stress, however, that we are still making a very important assumption: 
namely, that the cusp ellipticals present {\em today}, in a given stellar mass bin, 
are representative of the true progenitors of the core ellipticals present today at 
similar stellar mass. This is not necessarily the case! The progenitor galaxies 
had to form earlier than their descendants (and the stellar populations of the 
core population appear older, on average, than those of the cusp population). 
So it is not hard to imagine, for example, that the core progenitors formed from 
most gas-rich events at high redshift, and/or that they were more concentrated and 
compact, as observed spheroids at these redshifts 
\citep[e.g.][and references therein]{williams:2009.size.evol.disks.bulges.to.z2}. 
And some of the cusp population at high masses may represent cored galaxies that 
had their cusps ``rebuilt'' by new inflows from subsequent gas-rich mergers 
and/or cooling flows.
Indeed, structural evolution in the outer parts of these galaxies, 
and subsequent mergers, are expected 
in most theoretical models \citep{hopkins:r.z.evol,naab:size.evol.from.minor.mergers}
It does appear, however, that the central portions from $\sim1-5\,$kpc of even the most compact 
$z>2$ spheroids are similar in their stellar mass densities and profile 
shapes to local massive spheroids \citep{hopkins:density.galcores}, 
so our implicit assumption may be justified. Nevertheless, an ideal comparison would 
contrast present-day cored spheroids with higher-redshift candidate progenitors, chosen 
on the basis of a combination of mass and clustering or other constraints such that they 
are likely to evolve into the present-day core population. 
Unfortunately, there are no 
present observations that can resolve the cusp/core structure of these high-redshift 
systems. Therefore we are limited to the present day cusp population as our best-guess 
reference set, and should be careful to interpret our results in this context. 

With that caveat in mind, we show that there is a clear trend in the 
comparison of present-day nuclear profiles: at small radii, the distribution of 
enclosed mass in chosen cusp-core samples is bimodal at high significance. 
The mean mass difference -- the 
average ``scoured'' mass, in the scouring interpretation -- between 
the two bimodal peaks is low at small radii ($\ll M_{\rm BH}$ at 
$R \sim 3-10\,$pc), reflecting the fact that there is simply little absolute mass 
in such a small enclosed volume. 
The apparent ``scoured'' mass then grows 
with radius (roughly as $\propto R^{1.3-1.8}$) until $\sim 10-100$\,pc, at which 
point it asymptotes to a constant mass fraction $\approx 1-2\,M_{\rm BH}$ 
($\sim 10^{-3}-10^{-2.5}\,M_{\ast}$). The fact that we see this 
``turnover'' or asymptotic behavior suggests that we are actually resolving the 
{\em total} scoured mass, and radii affected. At larger 
radii $\gg 100\,$pc, there is no statistically significant signal for bimodality, 
and the $\sim 90\%$ upper limits on the scoured mass remain consistent 
with the same asymptotic mass difference. 
Neither the scoured mass fraction nor characteristic radii appear to 
scale strongly with galaxy mass; however, our dynamic range is limited to the mass range 
over which both cusp and core populations exist in significant 
numbers ($\sim10^{10.5-11.5}\,\msun$). 
The characteristic scouring radius, for example, may well scale with the BH radius of influence 
$R_{\rm BH}\sim G\,M_{\rm BH}\,\sigma^{-2} \propto M_{\ast}^{0.5}$, 
and in Figures~\ref{fig:fits}-\ref{fig:sigma}, there is a tentative hint 
of this (it appears that the characteristic radius increases with mass), 
but the difference is not statistically significant over this dynamic range 
with the present statistics. 

These results are consistent with the expectation from $N$-body experiments 
of the effects of a binary BH on the central stellar mass distribution. 
Such experiments typically find that the stellar mass scattered before the 
binary BH merges is $\sim 0.5-1.5\,M_{\rm BH}$ 
\citep[with dependence in that range on the central structural properties 
of the galaxy and structure of the 
inspiralling mass; see][and references therein]{merritt:mass.deficit,
perets:massive.perturber.bh.mgr}. 
The characteristic radii are also consistent; 
such simulations typically find that although stars from larger radii can 
contribute significantly to BH coalescence if they are on long radial box orbits in a 
triaxial potential \citep{quinlan:bh.binary.tang.orbit.bias} 
or if there is a strong perturber 
\citep{perets:binary.bh.loss.cone.filled.by.massive.perturbers}, 
a significant impact on the mass profile is restricted to
radii where the initial $M_{\ast}(<R)\lesssim M_{\rm BH}$; for a \citet{hernquist:profile} 
or $r^{1/4}$ bulge, this is approximately below 
$R/R_{\rm eff} \approx 0.5\,\sqrt{M_{\rm BH}/M_{ast}}$ or $\approx100\,$pc for an 
$\sim L_{\ast}$ bulge. 

Comparing this to studies where the mass ``deficit'' $\Delta M_{\ast}$ is determined 
from profile fitting, we find a smaller $\Delta M_{\ast}$ (and 
smaller scoured radii) than some. 
\citet{ferrarese:profiles} and \citet{kormendy:core.mass.deficits} 
estimate $\Delta M_{\ast}$ in a sub-set of the galaxies here, by comparing 
the actual light profile to the inward extrapolation of Sersic functions fitted to larger radii, 
they find $\Delta M_{\ast} \sim 2-4\,M_{\rm BH}$ and $\sim (10- 20) \,M_{\rm BH}$, respectively. 
However, we show in \S~\ref{sec:inferred} that these differences can arise 
as a systematic effect of the assumed profile shape. 
Fundamentally, at present, any such choice is somewhat arbitrary. 
And by assuming a specific 
functional form, the fits implicitly couple the assumed ``progenitor'' profile at 
small radii to the profile at large radii; for the Sersic fit, a galaxy with a more extended 
envelope (higher $n_{s}$) will necessarily imply a more steep nuclear 
Sersic profile extrapolation. Thus, for the {\em same} true nuclear profile, the 
inferred mass deficit and core radius increase. 
And in fact, in both samples, 
the observed core ellipticals have more extended envelopes and higher $n_{s}$ 
than essentially any of the cusp elliptical population -- thus the inward extrapolation of 
their Sersic profiles is not necessarily 
identical to the typical nuclear shape in cusps. 

\citet{lauer:massive.bhs} and \citet{milosavljevic:core.mass} 
consider instead the absolute mass 
enclosed within the radius where the profile steepens beyond a specified 
logarithmic slope (${\rm d}\ln I_{\nu}/{\rm d}\ln r = -0.5$)
or relative to the inwards extrapolation of a power-law (as opposed 
to Sersic) fit. They find $\Delta M_{\ast}\approx2.5\,M_{\rm BH}$ 
and $\sim0.5-10\,M_{\rm BH}$, respectively. 
Both cases, however, still require some implicit {\em a priori} assumption of a 
``correct'' functional form for the progenitor light profile at small radii 
(or the characteristic shape of ``scoured'' regions).
Moreover, in this approach, the mass measured is 
really the post-scouring mass inside this radius, not necessarily the mass removed 
(a non-trivial difference at the factor $\sim2-4$ level). 

In addition, the nature of the relatively flat profile at small radii means that 
whenever some functional form for the progenitor light profile is assumed, 
most of the ``missing light'' will come from the radii very near where the core 
breaks downwards relative to the ``non-core'' profile. 
As such, the inferred mass deficits are quite sensitive to the exact 
parameterization of the extrapolated profile near this radius (or, for a fixed 
slope cutoff, to the exact slope cut chosen). 
For the core-Sersic law, for example, 
there is a degeneracy between 
the fitted core break steepness parameter and the 
inner profile slope; small variations and common choices in fixing or freeing 
one of those parameters systematically affect the inferred mass deficits and 
radii. Our approach avoids this degeneracy; however, it remains 
difficult to statistically determine whether scouring has taken place 
at large radii where the enclosed mass is much larger than $M_{\rm BH}$. 

That being said, we do find that the median trend of mass deficit with 
radius, and absolute value inferred as a function of mass, can be 
approximately reproduced with the simple assumption that 
the ``progenitor'' profile represents a power-law continuation of 
the nuclear slope with some minimum ${\rm d}\ln I_{\nu}/{\rm d}\ln r \approx -0.6$, 
at radii where the slope flattens below this value. 
This is, of course, an assumption of the sort we are trying to avoid, but 
it does allow for estimation of the mass ``deficit'' on an individual galaxy basis, 
and our comparison here provides at least some justification for such a methodology. 
This also explains why our results are similar to those in 
\citet{lauer:massive.bhs} and \citet{milosavljevic:core.mass}. 
It would be of considerable physical interest to see whether or not there 
is any physical motivation for such a ``characteristic'' slope -- 
for example, it has been suggested that self-regulating inflow processes 
might set a characteristic slope around this value, in inflows driven 
by gravitational instabilities \citep{hopkins:zoom.sims,hopkins:m31.disk}. 

We stress that we agree with essentially all of the 
{\em qualitative} conclusions of the works discussed above. 
However, the resulting {\em quantitative} distinctions and possible 
systematics are important as tests of the theoretical models 
for core creation via BH binary mergers; very large $\Delta M_{\ast}$ and/or 
apparent core radii $\sim$kpc are actually quite difficult for models of 
core scouring to explain. 
Our more conservative study here therefore presents a natural and 
important complement to these works. We robustly 
set limits on the median scouring mass 
as a function of galaxy mass and radius without reference to any 
specific functional form for profile fitting. Moreover, to 
the extent that these measurements can be improved, they 
hold the potential to constrain not just the total mass that has been 
affected by scouring, but actually the scouring ``kernel'' -- i.e.\ the 
magnitude of the effect as a function of radius, 
which can break a number of degeneracies between theoretical 
models \citep[see the discussion in][]{perets:massive.perturber.bh.mgr}.

Clearly, the very simple methodology presented here has significant 
limitations. We find even with a sample of $\sim200$ well-observed 
spheroids that the statistical significance of our results is limited by 
the number of observed systems, because we can only consider systems 
in a narrow stellar mass range, and must in that range have a significant 
population of both cusp and core systems. As such, 
improving these constraints will require extending these samples with 
more observations of rare low-mass 
core or high-mass cusp ellipticals. Increasing the number of systems 
will also extend the radial range over which constraints can be applied; 
at some radius $R$ with an average enclosed mass $M_{\ast}(<R)$, 
the expected fractional offset between populations will be 
$\sim M_{\rm BH}/M_{\ast}(<R)$ -- this must be $\gtrsim 1/\sqrt{N}$ 
for a significant detection of the residual effects of ``scouring'' at these radii. 
In particular, determining whether or not the scoured mass is really ejected to 
large radii (or unbound) or simply kicked to moderate $\sim0.1-1\,$kpc radii 
will require extended samples along these lines. With such observations, 
study of the ``scouring profile'' ($\Delta\,M_{\ast}(<R)$) 
can be used to determine the differential contribution to scouring/BH 
binary coalescence from stars at different radii. This will allow direct observational 
probes of the loss cone, in particular of which stellar radii and orbits contribute. 

The methodology here also only constrains the median mass difference 
between populations -- it cannot be used on an object-by-object basis. But 
with significantly larger samples, such that not just the median but the scatter 
and shape of each distribution $M_{\ast}(<R)$ can be determined, 
then the variation in scouring effects can be constrained (at least in a 
statistical sense). 

Independent tests are also important. For example, 
scouring is expected to preferentially eliminate stars 
on radial orbits and leave a bias for tangential orbits 
\citep[][]{quinlan:bh.binary.tang.orbit.bias}. \citet{gebhardt:nuclear.anisotropies} see 
tentative evidence for this in a limited sample of ellipticals; the 
major-axis radii within which the effect appears are generally 
$\sim0.5-3\,R_{\rm BH}$, similar to our estimates here. 
Ultimately, a combination of these independent constraints and the 
non-parametric constraints presented here can be used to determined 
whether or not, and if so which, parametric approach(es) are most faithful to 
the true scoured mass, and motivate their use on an object-by-object basis.

\acknowledgments 
We thank Hagai Perets for insightful comments in the development of 
this manuscript. We also thank John Kormendy, Tod Lauer, Eliot Quataert, 
and Norm Murray for helpful discussions at various stages in this and related 
work. Support for PFH was provided by the Miller Institute for Basic Research 
in Science, University of California Berkeley.
EQ is supported in part by NASA grant NNG06GI68G and 
the David and Lucile Packard Foundation.
\\

\bibliography{/Users/phopkins/Documents/lars_galaxies/papers/ms}

\begin{thebibliography}{92}
\expandafter\ifx\csname natexlab\endcsname\relax\def\natexlab#1{#1}\fi

\bibitem[{{Aller} \& {Richstone}(2007)}]{aller:mbh.esph}
{Aller}, M.~C., \& {Richstone}, D.~O. 2007, \apj, 665, 120

\bibitem[{{Barnes}(1988)}]{barnes:disk.halo.mergers}
{Barnes}, J.~E. 1988, \apj, 331, 699

\bibitem[{{Begelman} {et~al.}(1980){Begelman}, {Blandford}, \&
  {Rees}}]{begelman:scouring}
{Begelman}, M.~C., {Blandford}, R.~D., \& {Rees}, M.~J. 1980, \nat, 287, 307

\bibitem[{{Bell} {et~al.}(2003){Bell}, {McIntosh}, {Katz}, \&
  {Weinberg}}]{bell:mfs}
{Bell}, E.~F., {McIntosh}, D.~H., {Katz}, N., \& {Weinberg}, M.~D. 2003, \apjs,
  149, 289

\bibitem[{{Bender}(1988)}]{bender:88.shapes}
{Bender}, R. 1988, \aap, 193, L7

\bibitem[{{Bender} {et~al.}(1992){Bender}, {Burstein}, \&
  {Faber}}]{bender:ell.kinematics}
{Bender}, R., {Burstein}, D., \& {Faber}, S.~M. 1992, \apj, 399, 462

\bibitem[{{Bender} {et~al.}(1989){Bender}, {Surma}, {Doebereiner},
  {Moellenhoff}, \& {Madejsky}}]{bender89}
{Bender}, R., {Surma}, P., {Doebereiner}, S., {Moellenhoff}, C., \& {Madejsky},
  R. 1989, \aap, 217, 35

\bibitem[{{Berczik} {et~al.}(2006){Berczik}, {Merritt}, {Spurzem}, \&
  {Bischof}}]{berczik:triaxial.bh.mergers}
{Berczik}, P., {Merritt}, D., {Spurzem}, R., \& {Bischof}, H.-P. 2006, \apjl,
  642, L21

\bibitem[{{Bezanson} {et~al.}(2009){Bezanson}, {van Dokkum}, {Tal},
  {Marchesini}, {Kriek}, {Franx}, \& {Coppi}}]{bezanson:massive.gal.cores.evol}
{Bezanson}, R., {van Dokkum}, P.~G., {Tal}, T., {Marchesini}, D., {Kriek}, M.,
  {Franx}, M., \& {Coppi}, P. 2009, \apj, 697, 1290

\bibitem[{{Boylan-Kolchin} {et~al.}(2005){Boylan-Kolchin}, {Ma}, \&
  {Quataert}}]{boylankolchin:mergers.fp}
{Boylan-Kolchin}, M., {Ma}, C.-P., \& {Quataert}, E. 2005, \mnras, 362, 184

\bibitem[{{Boylan-Kolchin} {et~al.}(2006){Boylan-Kolchin}, {Ma}, \&
  {Quataert}}]{boylankolchin:dry.mergers}
---. 2006, \mnras, 369, 1081

\bibitem[{{Bracewell}(1965)}]{bracewell:book}
{Bracewell}, R. 1965, {The Fourier Transform and its applications} (McGraw-Hill
  Electrical and Electronic Engineering Series, New York: McGraw-Hill)

\bibitem[{{Cappellari} {et~al.}(2007)}]{cappellari:anisotropy}
{Cappellari}, M., {et~al.} 2007, \mnras, 379, 418

\bibitem[{{Chabrier}(2003)}]{chabrier:imf}
{Chabrier}, G. 2003, \pasp, 115, 763

\bibitem[{{C{\^o}t{\'e}} {et~al.}(2006)}]{cote:virgo}
{C{\^o}t{\'e}}, P., {et~al.} 2006, \apjs, 165, 57

\bibitem[{{C{\^o}t{\'e}} {et~al.}(2007)}]{cote:smooth.transition}
---. 2007, \apj, 671, 1456

\bibitem[{{Crane} {et~al.}(1993)}]{crane93}
{Crane}, P., {et~al.} 1993, \aj, 106, 1371

\bibitem[{{Davies} {et~al.}(1983){Davies}, {Efstathiou}, {Fall}, {Illingworth},
  \& {Schechter}}]{davies:faint.ell.kinematics}
{Davies}, R.~L., {Efstathiou}, G., {Fall}, S.~M., {Illingworth}, G., \&
  {Schechter}, P.~L. 1983, \apj, 266, 41

\bibitem[{{Davis} {et~al.}(1985){Davis}, {Cawson}, {Davies}, \&
  {Illingworth}}]{davis:85}
{Davis}, L.~E., {Cawson}, M., {Davies}, R.~L., \& {Illingworth}, G. 1985, \aj,
  90, 169

\bibitem[{{Emsellem} {et~al.}(2004)}]{emsellem:sauron.rotation.data}
{Emsellem}, E., {et~al.} 2004, \mnras, 352, 721

\bibitem[{{Emsellem} {et~al.}(2007)}]{emsellem:sauron.rotation}
---. 2007, \mnras, 379, 401

\bibitem[{{Faber} {et~al.}(1997)}]{faber:ell.centers}
{Faber}, S.~M., {et~al.} 1997, \aj, 114, 1771

\bibitem[{{Ferrarese} \& {Merritt}(2000)}]{FM00}
{Ferrarese}, L., \& {Merritt}, D. 2000, \apjl, 539, L9

\bibitem[{{Ferrarese} {et~al.}(1994){Ferrarese}, {van den Bosch}, {Ford},
  {Jaffe}, \& {O'Connell}}]{ferrarese:type12}
{Ferrarese}, L., {van den Bosch}, F.~C., {Ford}, H.~C., {Jaffe}, W., \&
  {O'Connell}, R.~W. 1994, \aj, 108, 1598

\bibitem[{{Ferrarese} {et~al.}(2006)}]{ferrarese:profiles}
{Ferrarese}, L., {et~al.} 2006, \apjs, 164, 334

\bibitem[{{Gallagher} \& {Ostriker}(1972)}]{gallagherostriker72}
{Gallagher}, III, J.~S., \& {Ostriker}, J.~P. 1972, \aj, 77, 288

\bibitem[{{Gebhardt} {et~al.}(1996)}]{gebhardt96}
{Gebhardt}, K., {et~al.} 1996, \aj, 112, 105

\bibitem[{{Gebhardt} {et~al.}(2000)}]{Gebhardt00}
---. 2000, \apjl, 539, L13

\bibitem[{{Gebhardt} {et~al.}(2003)}]{gebhardt:nuclear.anisotropies}
---. 2003, \apj, 583, 92

\bibitem[{{Graham} {et~al.}(2001){Graham}, {Erwin}, {Caon}, \&
  {Trujillo}}]{graham:concentration}
{Graham}, A.~W., {Erwin}, P., {Caon}, N., \& {Trujillo}, I. 2001, \apjl, 563,
  L11

\bibitem[{{Graham} {et~al.}(2003){Graham}, {Erwin}, {Trujillo}, \& {Asensio
  Ramos}}]{graham:core.sersic}
{Graham}, A.~W., {Erwin}, P., {Trujillo}, I., \& {Asensio Ramos}, A. 2003, \aj,
  125, 2951

\bibitem[{{Gualandris} \& {Merritt}(2007)}]{gualandrismerritt:scouring.review}
{Gualandris}, A., \& {Merritt}, D. 2007, in Black Holes, ed. M.~{Livio} \&
  A.~M. {Koekemoer}, STScI Spring Symposium [arXiv:0708.3083]

\bibitem[{{H{\"a}ring} \& {Rix}(2004)}]{haringrix}
{H{\"a}ring}, N., \& {Rix}, H.-W. 2004, \apjl, 604, L89

\bibitem[{{Hartigan} \& {Hartigan}(1985)}]{hartigan:dip.bimodality.test}
{Hartigan}, J.~A., \& {Hartigan}, P.~M. 1985, The Annals of Statistics, 13, 70

\bibitem[{{Hernquist}(1990)}]{hernquist:profile}
{Hernquist}, L. 1990, \apj, 356, 359

\bibitem[{{Hernquist}(1992)}]{hernquist:bulgeless.mergers}
---. 1992, \apj, 400, 460

\bibitem[{{Hernquist}(1993)}]{hernquist:bulge.mergers}
---. 1993, \apj, 409, 548

\bibitem[{{Hernquist} {et~al.}(1993){Hernquist}, {Spergel}, \&
  {Heyl}}]{hernquist:phasespace}
{Hernquist}, L., {Spergel}, D.~N., \& {Heyl}, J.~S. 1993, \apj, 416, 415

\bibitem[{{Hibbard} \& {Yun}(1999)}]{hibbard.yun:excess.light}
{Hibbard}, J.~E., \& {Yun}, M.~S. 1999, \apjl, 522, L93

\bibitem[{{Holley-Bockelmann} \&
  {Sigurdsson}(2006)}]{holley:triaxial.loss.cone}
{Holley-Bockelmann}, K., \& {Sigurdsson}, S. 2006, \mnras, in press,
  arXiv:astro-ph/0601520

\bibitem[{{Hopkins} {et~al.}(2010){Hopkins}, {Bundy}, {Hernquist}, {Wuyts}, \&
  {Cox}}]{hopkins:r.z.evol}
{Hopkins}, P.~F., {Bundy}, K., {Hernquist}, L., {Wuyts}, S., \& {Cox}, T.~J.
  2010, \mnras, 401, 1099

\bibitem[{{Hopkins} {et~al.}(2009{\natexlab{a}}){Hopkins}, {Bundy}, {Murray},
  {Quataert}, {Lauer}, \& {Ma}}]{hopkins:density.galcores}
{Hopkins}, P.~F., {Bundy}, K., {Murray}, N., {Quataert}, E., {Lauer}, T.~R., \&
  {Ma}, C.-P. 2009{\natexlab{a}}, \mnras, 398, 898

\bibitem[{{Hopkins} {et~al.}(2009{\natexlab{b}}){Hopkins}, {Cox}, {Dutta},
  {Hernquist}, {Kormendy}, \& {Lauer}}]{hopkins:cusps.ell}
{Hopkins}, P.~F., {Cox}, T.~J., {Dutta}, S.~N., {Hernquist}, L., {Kormendy},
  J., \& {Lauer}, T.~R. 2009{\natexlab{b}}, \apjs, 181, 135

\bibitem[{{Hopkins} {et~al.}(2008{\natexlab{a}}){Hopkins}, {Cox}, \&
  {Hernquist}}]{hopkins:cusps.fp}
{Hopkins}, P.~F., {Cox}, T.~J., \& {Hernquist}, L. 2008{\natexlab{a}}, \apj,
  689, 17

\bibitem[{{Hopkins} {et~al.}(2008{\natexlab{b}}){Hopkins}, {Hernquist}, {Cox},
  {Dutta}, \& {Rothberg}}]{hopkins:cusps.mergers}
{Hopkins}, P.~F., {Hernquist}, L., {Cox}, T.~J., {Dutta}, S.~N., \& {Rothberg},
  B. 2008{\natexlab{b}}, \apj, 679, 156

\bibitem[{{Hopkins} {et~al.}(2009{\natexlab{c}}){Hopkins}, {Hernquist}, {Cox},
  {Kere{\v s}}, \& {Wuyts}}]{hopkins:cusps.evol}
{Hopkins}, P.~F., {Hernquist}, L., {Cox}, T.~J., {Kere{\v s}}, D., \& {Wuyts},
  S. 2009{\natexlab{c}}, \apj, 691, 1424

\bibitem[{{Hopkins} {et~al.}(2007{\natexlab{a}}){Hopkins}, {Hernquist}, {Cox},
  {Robertson}, \& {Krause}}]{hopkins:bhfp.theory}
{Hopkins}, P.~F., {Hernquist}, L., {Cox}, T.~J., {Robertson}, B., \& {Krause},
  E. 2007{\natexlab{a}}, \apj, 669, 45

\bibitem[{{Hopkins} {et~al.}(2007{\natexlab{b}}){Hopkins}, {Hernquist}, {Cox},
  {Robertson}, \& {Krause}}]{hopkins:bhfp.obs}
---. 2007{\natexlab{b}}, \apj, 669, 67

\bibitem[{{Hopkins} {et~al.}(2009{\natexlab{d}}){Hopkins}, {Lauer}, {Cox},
  {Hernquist}, \& {Kormendy}}]{hopkins:cores}
{Hopkins}, P.~F., {Lauer}, T.~R., {Cox}, T.~J., {Hernquist}, L., \& {Kormendy},
  J. 2009{\natexlab{d}}, \apjs, 181, 486

\bibitem[{{Hopkins} {et~al.}(2009{\natexlab{e}}){Hopkins}, {Murray}, \&
  {Thompson}}]{hopkins:msigma.scatter}
{Hopkins}, P.~F., {Murray}, N., \& {Thompson}, T.~A. 2009{\natexlab{e}},
  \mnras, 398, 303

\bibitem[{{Hopkins} \& {Quataert}(2009)}]{hopkins:zoom.sims}
{Hopkins}, P.~F., \& {Quataert}, E. 2009, \mnras, in press, arXiv:0912.3257

\bibitem[{{Hopkins} \& {Quataert}(2010)}]{hopkins:m31.disk}
---. 2010, \mnras, L62+

\bibitem[{{Hyde} \& {Bernardi}(2008)}]{hyde:stellar.mass.fp}
{Hyde}, J.~B., \& {Bernardi}, M. 2008, \mnras, in press, arXiv:0810.4924
  [astro-ph]

\bibitem[{{Jedrzejewski} {et~al.}(1987){Jedrzejewski}, {Davies}, \&
  {Illingworth}}]{jedrzejewski:87}
{Jedrzejewski}, R.~I., {Davies}, R.~L., \& {Illingworth}, G.~D. 1987, \aj, 94,
  1508

\bibitem[{{King}(1978)}]{king78}
{King}, I.~R. 1978, \apj, 222, 1

\bibitem[{{Kormendy}(1985)}]{kormendy85:profiles}
{Kormendy}, J. 1985, \apjl, 292, L9

\bibitem[{{Kormendy}(1987)}]{kormendy:cores.review}
{Kormendy}, J. 1987, in IAU Symposium, Vol. 127, Structure and Dynamics of
  Elliptical Galaxies (Dordrecht, D. Reidel Publishing Co.), ed. P.~T. {de
  Zeeuw}, 17--34

\bibitem[{{Kormendy}(1999)}]{kormendy99}
{Kormendy}, J. 1999, in Astronomical Society of the Pacific Conference Series,
  Vol. 182, Galaxy Dynamics - A Rutgers Symposium, ed. D.~R. {Merritt},
  M.~{Valluri}, \& J.~A. {Sellwood}, 124--+

\bibitem[{{Kormendy} \& {Bender}(1996)}]{kormendybender96}
{Kormendy}, J., \& {Bender}, R. 1996, \apjl, 464, L119+

\bibitem[{{Kormendy} \& {Bender}(2009)}]{kormendy:core.mass.deficits}
---. 2009, \apjl, 691, L142

\bibitem[{{Kormendy} {et~al.}(2009){Kormendy}, {Fisher}, {Cornell}, \&
  {Bender}}]{jk:profiles}
{Kormendy}, J., {Fisher}, D.~B., {Cornell}, M.~E., \& {Bender}, R. 2009, \apjs,
  182, 216

\bibitem[{{Kormendy} \& {Illingworth}(1982)}]{kormendy:bulge.rotation}
{Kormendy}, J., \& {Illingworth}, G. 1982, \apj, 256, 460

\bibitem[{{Laine} {et~al.}(2003){Laine}, {van der Marel}, {Lauer}, {Postman},
  {O'Dea}, \& {Owen}}]{laine:03}
{Laine}, S., {van der Marel}, R.~P., {Lauer}, T.~R., {Postman}, M., {O'Dea},
  C.~P., \& {Owen}, F.~N. 2003, \aj, 125, 478

\bibitem[{{Lauer}(1985)}]{lauer85:cores}
{Lauer}, T.~R. 1985, \apj, 292, 104

\bibitem[{{Lauer} {et~al.}(1995){Lauer}, {Ajhar}, {Byun}, {Dressler}, {Faber},
  {Grillmair}, {Kormendy}, {Richstone}, \& {Tremaine}}]{lauer:95}
{Lauer}, T.~R., {Ajhar}, E.~A., {Byun}, Y.-I., {Dressler}, A., {Faber}, S.~M.,
  {Grillmair}, C., {Kormendy}, J., {Richstone}, D., \& {Tremaine}, S. 1995,
  \aj, 110, 2622

\bibitem[{{Lauer} {et~al.}(1992)}]{lauer92}
{Lauer}, T.~R., {et~al.} 1992, \aj, 103, 703

\bibitem[{{Lauer} {et~al.}(2005)}]{lauer:centers}
---. 2005, \aj, 129, 2138

\bibitem[{{Lauer} {et~al.}(2007{\natexlab{a}})}]{lauer:bimodal.profiles}
---. 2007{\natexlab{a}}, \apj, 664, 226

\bibitem[{{Lauer} {et~al.}(2007{\natexlab{b}})}]{lauer:massive.bhs}
---. 2007{\natexlab{b}}, \apj, 662, 808

\bibitem[{{Magorrian} {et~al.}(1998)}]{magorrian}
{Magorrian}, J., {et~al.} 1998, \aj, 115, 2285

\bibitem[{{Marconi} \& {Hunt}(2003)}]{marconihunt}
{Marconi}, A., \& {Hunt}, L.~K. 2003, \apjl, 589, L21

\bibitem[{{McDermid} {et~al.}(2006)}]{mcdermid:sauron.profiles}
{McDermid}, R.~M., {et~al.} 2006, \mnras, 373, 906

\bibitem[{{Merritt}(2006)}]{merritt:mass.deficit}
{Merritt}, D. 2006, \apj, 648, 976

\bibitem[{{Milosavljevi{\'c}} {et~al.}(2002){Milosavljevi{\'c}}, {Merritt},
  {Rest}, \& {van den Bosch}}]{milosavljevic:core.mass}
{Milosavljevi{\'c}}, M., {Merritt}, D., {Rest}, A., \& {van den Bosch}, F.~C.
  2002, \mnras, 331, L51

\bibitem[{{Naab} {et~al.}(2009){Naab}, {Johansson}, \&
  {Ostriker}}]{naab:size.evol.from.minor.mergers}
{Naab}, T., {Johansson}, P.~H., \& {Ostriker}, J.~P. 2009, \apjl, 699, L178

\bibitem[{{Peletier} {et~al.}(1990){Peletier}, {Davies}, {Illingworth},
  {Davis}, \& {Cawson}}]{peletier:profiles}
{Peletier}, R.~F., {Davies}, R.~L., {Illingworth}, G.~D., {Davis}, L.~E., \&
  {Cawson}, M. 1990, \aj, 100, 1091

\bibitem[{{Perets} \& {Alexander}(2008)}]{perets:massive.perturber.bh.mgr}
{Perets}, H.~B., \& {Alexander}, T. 2008, \apj, 677, 146

\bibitem[{{Perets} {et~al.}(2007){Perets}, {Hopman}, \&
  {Alexander}}]{perets:binary.bh.loss.cone.filled.by.massive.perturbers}
{Perets}, H.~B., {Hopman}, C., \& {Alexander}, T. 2007, \apj, 656, 709

\bibitem[{{Quillen} {et~al.}(2000){Quillen}, {Bower}, \&
  {Stritzinger}}]{quillen:00}
{Quillen}, A.~C., {Bower}, G.~A., \& {Stritzinger}, M. 2000, \apjs, 128, 85

\bibitem[{{Quinlan} \& {Hernquist}(1997)}]{quinlan:bh.binary.tang.orbit.bias}
{Quinlan}, G.~D., \& {Hernquist}, L. 1997, New Astronomy, 2, 533

\bibitem[{{Ravindranath} {et~al.}(2001){Ravindranath}, {Ho}, {Peng},
  {Filippenko}, \& {Sargent}}]{ravindranath:01}
{Ravindranath}, S., {Ho}, L.~C., {Peng}, C.~Y., {Filippenko}, A.~V., \&
  {Sargent}, W.~L.~W. 2001, \aj, 122, 653

\bibitem[{{Rest} {et~al.}(2001){Rest}, {van den Bosch}, {Jaffe}, {Tran},
  {Tsvetanov}, {Ford}, {Davies}, \& {Schafer}}]{rest:01}
{Rest}, A., {van den Bosch}, F.~C., {Jaffe}, W., {Tran}, H., {Tsvetanov}, Z.,
  {Ford}, H.~C., {Davies}, J., \& {Schafer}, J. 2001, \aj, 121, 2431

\bibitem[{{Rothberg} \& {Joseph}(2004)}]{rj:profiles}
{Rothberg}, B., \& {Joseph}, R.~D. 2004, \aj, 128, 2098

\bibitem[{{Sesana} {et~al.}(2007){Sesana}, {Haardt}, \&
  {Madau}}]{sesana:binary.bh.mergers}
{Sesana}, A., {Haardt}, F., \& {Madau}, P. 2007, \apj, 660, 546

\bibitem[{{Simien} \& {Prugniel}(1997{\natexlab{a}})}]{simien:kinematics.1}
{Simien}, F., \& {Prugniel}, P. 1997{\natexlab{a}}, \aaps, 122, 521

\bibitem[{{Simien} \& {Prugniel}(1997{\natexlab{b}})}]{simien:kinematics.2}
---. 1997{\natexlab{b}}, \aaps, 126, 15

\bibitem[{{Simien} \& {Prugniel}(1997{\natexlab{c}})}]{simien:kinematics.3}
---. 1997{\natexlab{c}}, \aaps, 126, 519

\bibitem[{{Tremaine} {et~al.}(2002)}]{tremaine:msigma}
{Tremaine}, S., {et~al.} 2002, \apj, 574, 740

\bibitem[{{Trujillo} {et~al.}(2002){Trujillo}, {Asensio Ramos},
  {Rubi{\~n}o-Mart{\'{\i}}n}, {Graham}, {Aguerri}, {Cepa}, \&
  {Guti{\'e}rrez}}]{trujillo:sersic.fits}
{Trujillo}, I., {Asensio Ramos}, A., {Rubi{\~n}o-Mart{\'{\i}}n}, J.~A.,
  {Graham}, A.~W., {Aguerri}, J.~A.~L., {Cepa}, J., \& {Guti{\'e}rrez}, C.~M.
  2002, \mnras, 333, 510

\bibitem[{{Williams} {et~al.}(2010){Williams}, {Quadri}, {Franx}, {van Dokkum},
  {Toft}, {Kriek}, \& {Labb{\'e}}}]{williams:2009.size.evol.disks.bulges.to.z2}
{Williams}, R.~J., {Quadri}, R.~F., {Franx}, M., {van Dokkum}, P., {Toft}, S.,
  {Kriek}, M., \& {Labb{\'e}}, I. 2010, \apj, 713, 738

\bibitem[{{Young} {et~al.}(1978){Young}, {Westphal}, {Kristian}, {Wilson}, \&
  {Landauer}}]{young78}
{Young}, P.~J., {Westphal}, J.~A., {Kristian}, J., {Wilson}, C.~P., \&
  {Landauer}, F.~P. 1978, \apj, 221, 721

\bibitem[{{Zier}(2007)}]{zier:binary.bh.loss.cone.dynamics}
{Zier}, C. 2007, \mnras, 378, 1309

\end{thebibliography}

\end{document}